\documentstyle[epsfig,12pt]{article}

\renewcommand{\theequation}{\mbox{$\thesection . \arabic{equation}$}}

\begin{document}
\newcommand{\beq}{\begin{equation}}
\newcommand{\eeq}{\end{equation}}
\newcommand{\barr}{\begin{eqnarray}}
\newcommand{\earr}{\end{eqnarray}}

\newcommand{\andy}[1]{ }

\def\rz{\mbox{\boldmath $y$}_0}
\def\a{\alpha_0} \def\da{\delta\alpha}
\def\Da{D_\alpha}
\def\h{\widehat}
\def\t{\widetilde}
\def\cH{{\cal H}}
\def\ud{\uparrow\downarrow}
\def\uu{\uparrow}
\def\dd{\downarrow}
\def\u{\uparrow}
\def\d{\downarrow}
\def\b1{{\bf 1}}

\def\coltwovector#1#2{\left({#1\atop#2}\right)}
\def\upp{\coltwovector10}    \def\downn{\coltwovector01}


\def\ask{\marginpar{?? ask:  \hfill}}
\def\fin{\marginpar{fill in ... \hfill}}
\def\note{\marginpar{note \hfill}}
\def\check{\marginpar{check \hfill}}
\def\discuss{\marginpar{discuss \hfill}}

\begin{titlepage}
\begin{flushright}
\today \\
BA-TH/316-99\\ WU-HEP-99-1\\
\end{flushright}
\vspace{.5cm}
\begin{center}
{\Large\bf Reflection and Transmission in a Neutron-Spin Test of the
Quantum Zeno Effect} \\ [.5cm]

{Ken Machida,$^{(1)}$ Hiromichi Nakazato,$^{(1)}$ Saverio
Pascazio,$^{(2)}$ \\
        Helmut Rauch$^{(3)}$ and Sixia Yu$^{(4)}$ \\
           \quad    \\}
{\small\it  $^{(1)}$Department of Physics, Waseda University
                Tokyo 169-8555, Japan \\
        $^{(2)}$Dipartimento di Fisica, Universit\^^ {a} di Bari
                and
                Istituto Nazionale di Fisica Nucleare, Sezione di Bari,
                I-70126  Bari, Italy  \\
        $^{(3)}$Atominstitut der \"Osterreichischen Universit\"aten\\
                Stadionallee 2, A-1020, Wien, Austria \\
        $^{(4)}$Institute of Theoretical Physics, Academia Sinica,
                Beijing 100080, P.R.\ China\\
}
\vspace*{1cm}

{\small\bf Abstract} \\ \end{center}

{\small
The dynamics of a quantum system undergoing frequent ``measurements,"
leading to the so-called quantum Zeno effect, is examined on the basis of a
neutron-spin experiment recently proposed for its demonstration.
When the spatial degrees of freedom are duely taken
into account, neutron-reflection effects become very important and may
lead to an evolution which is totally different from the ideal case.

\vspace*{.5cm}
}

\end{titlepage}

\newpage

\setcounter{equation}{0}
\section{Introduction  }
\label{sec-intro}
\andy{intro}

A quantum system, prepared in a state that does not belong to an eigenvalue
of the total Hamiltonian, starts to evolve
quadratically in time \cite{Beskow,Misra}.
This characteristic behavior leads to the so-called quantum Zeno phenomenon,
namely the possibility of slowing down the temporal
evolution (eventually hindering
transitions to states different from the initial one) \cite{strev}.

The original proposals that aimed at verifying this effect
involved unstable systems and were not amenable to
experimental test \cite{Wilkinson}. However, the remarkable idea
\cite{Cook} to use a two-level system motivated an interesting
experimental test \cite{Itano},
revitalizing a debate on the physical meaning of this
phenomenon \cite{Itanodiscuss,PNBR}.
There seem to be a certain consensus, nowadays, that the quantum Zeno effect
(QZE) can be given a dynamical explanation, involving only an
explicit Hamiltonian dynamics.

It is worth emphasizing that the discussion of the last
few years mostly stemmed from experimental considerations, related to the
{\em practical} possibility of performing experimental tests. Some examples are
the interesting issue of ``interaction-free" measurements \cite{inn} and the
neutron-spin tests of the QZE \cite{PNBR,NNPR}.
In practical cases, one cannot neglect the presence of losses and
imperfections, which obviously conspire against an almost-ideal
experimental realization, more so when the total number of ``measurements"
increases above certain theoretical limits.

The aim of the present paper is to investigate an interesting (and often
overlooked) feature of what we might call the quantum Zeno dynamics. We shall
see that a series of ``measurements" (von Neumann's projections \cite{von})
does not necessarily hinder the evolution of the quantum system. On the
contrary, the system can evolve away from its initial state, provided it
remains in the subspace defined by the ``measurement" itself. This interesting
feature is readily understandable in terms of rigorous theorems
\cite{Misra}, but it seems to us that it is worth clarifying it by
analyzing interesting physical examples. We shall therefore focus our attention
on an experiment involving neutron spin \cite{PNBR}
and shall see that in fact this enables
us to kill two birds with one stone: not only the state of the neutron
undergoing QZE {\em will change}, but it will do so in a way that clarifies
why reflection effects may play a substantial role in the experiment analyzed.

In the neutron-spin example to be considered, the evolution of the
spin state is hindered  when a series of spectral decompositions
(in Wigner's sense \cite{Wigner}) is performed on the spin state.
No ``observation" of the spin states, and therefore no projection
{\em \`a la} von Neumann is required, as far as the different
branch waves of the wave function cannot interfere after the
spectral decomposition. Needless to say, the analysis that follows
could be performed in terms of a Hamiltonian dynamics, without
making use of projection operators. However, we shall use in this
paper the von Neumann technique, which will be found convenient
because it sheds light on some remarkable aspects of the Zeno
phenomenon and helps to pin down the physical implications of some
mathematical hypotheses with relatively less efforts.

The paper is organized as follows. We briefly review, in the next
section, the seminal theorem for the short-time dynamics of quantum
systems, proved by Misra and Sudarshan \cite{Misra}. Its
application to the neutron-spin case is discussed in
Sec.~\ref{sec-neutr}. In Secs.~\ref{sec-neutrspin} and
\ref{sec-compltrans}, unlike in previous papers \cite{PNBR,NNPR},
we shall incorporate the spatial (1-dimensional, for simplicity)
degrees of freedom of the neutron and represent them by an
additional quantum number that labels, roughly speaking, the
direction of motion of the wave packet. A more realistic analysis
is presented in Sec.~\ref{sec-numan}. Finally,
Sec.~\ref{sec-findisc} is devoted to a discussion. Some additional
aspects of our analysis are clarified in the Appendix.

\setcounter{equation}{0}
\section{Misra and Sudarshan's theorem }
\label{sec-MisSud}
\andy{MisSud}

Consider a quantum system Q, whose
states belong to the Hilbert space ${\cal H}$ and whose evolution
is described by the unitary operator $U(t)=\exp(-iHt)$, where $H$
is a semi-bounded Hamiltonian. Let $E$ be a projection operator and
$E{\cal H}E={\cal H}_E$ the subspace spanned by its eigenstates.
The initial density matrix $\rho_0$ of system Q is taken to belong
to ${\cal H}_E$. If Q is let to follow its ``undisturbed"
evolution, under the action of the Hamiltonian $H$ (i.e., no
measurements are performed in order to get informations about its
quantum state), the final state at time $T$ reads
\andy{noproie}
\beq
\rho (T) = U(T) \rho_0 U^\dagger (T)
  \label{eq:noproie}
\eeq
and the probability that the system is still in ${\cal H}_E$ at time $T$
 is
\andy{stillun}
\beq
P(T) = \mbox{Tr} \left[ U(T) \rho_0 U^\dagger(T) E \right] .
\label{eq:stillun}
\eeq
We call this a ``survival probability:" it
is in general smaller than 1, since the Hamiltonian $H$
induces transitions out of ${\cal H}_E$.
We shall say that the quantum systems has ``survived" if it is
found to be in ${\cal H}_E$ by means of a suitable measurement
process \cite{MScomment}.

Assume that we perform a measurement at time $t$,
in order to check whether Q has survived. Such a measurement
is formally represented by the projection operator $E$. By definition,
\andy{inprep}
\beq
\rho_0 = E \rho_0 E , \qquad \mbox{Tr} [ \rho_0 E ] = 1 .
\label{eq:inprep}
\eeq
After the measurement, the state of Q changes into
\andy{proie}
\beq
\rho_0 \rightarrow \rho(t) = E U(t) \rho_0 U^\dagger(t) E/P(t),
\label{eq:proie}
\eeq
where
\andy{probini}
\beq
P(t) = \mbox{Tr} \left[ U(t) \rho_0 U^\dagger(t) E \right]
\label{eq:probini}
\eeq
is the probability that the system has survived. [There is, of
course, a probability $1-P$ that the system has not survived (i.e.,
it has made a transition outside ${\cal H}_E$) and its state has
changed into $\rho^\prime(t) = (1-E) U(t) \rho_0 U^\dagger(t)
(1-E)/(1-P)$. We concentrate henceforth our attention on the
measurement outcome (\ref{eq:proie})-(\ref{eq:probini}).] The above
is the standard Copenhagen interpretation: The measurement is
considered to be instantaneous. The ``quantum Zeno paradox"
\cite{Misra} is the following. We prepare Q in the initial state
$\rho_0$ at time 0 and perform a series of $E$-observations at
times $t_k=kT/N \; (k=1,
\cdots, N)$. The state of Q after the above-mentioned $N$
measurements reads
\andy{Nproie}
\beq
\rho^{(N)}(T) = V_N(T) \rho_0 V_N^\dagger(T) , \qquad
    V_N(T) \equiv [ E U(T/N) E ]^N
\label{eq:Nproie}
\eeq
and the probability to find the system in ${\cal H}_E$ (``survival
probability") is given by
\andy{probNob}
\beq
P^{(N)}(T) = \mbox{Tr} \left[ V_N(T) \rho_0 V_N^\dagger(T) \right].
\label{eq:probNob}
\eeq
Equations (\ref{eq:Nproie})-(\ref{eq:probNob}) display the
``quantum Zeno effect:" repeated
observations in succession modify the dynamics of the quantum system;
under general conditions, if $N$ is sufficiently large, all transitions
outside ${\cal H}_E$ are inhibited.

In order to consider the $N \rightarrow \infty$ limit (``continuous
observation"), one needs some mathematical requirements: define
\andy{slim}
\beq
{\cal V} (T) \equiv \lim_{N \rightarrow \infty} V_N(T) ,
  \label{eq:slim}
\eeq
provided the above limit exists in the strong sense.
The final state of Q is then
\andy{infproie}
\beq
\t{\rho} (T) = {\cal V}(T) \rho_0 {\cal V}^\dagger (T)
  \label{eq:infproie}
\eeq
and the probability to find the system in ${\cal H}_E$ is
\andy{probinfob}
\beq
{\cal P} (T) \equiv \lim_{N \rightarrow \infty} P^{(N)}(T)
   = \mbox{Tr} \left[ {\cal V}(T) \rho_0 {\cal V}^\dagger(T) \right].
\label{eq:probinfob}
\eeq
One should carefully notice that nothing is said about the final
state $\t{\rho} (T)$, which depends on the characteristics of the
model investigated and on the {\em very measurement performed}
(i.e.\ on the projection operator $E$, which enters in the
definition of $V_N$). Misra and Sudarshan assumed, on physical
grounds, the strong continuity of ${\cal V}(t)$:
\andy{phgr}
\beq
\lim_{t \rightarrow 0^+} {\cal V}(t) = E
\label{eq:phgr}
\eeq
and proved that under general conditions the operators ${\cal
V}(T)$ (exist for all real $T$ and) form a semigroup labeled by the
time parameter $T$. Moreover, ${\cal V}^\dagger (T) = {\cal
V}(-T)$, so that ${\cal V}^\dagger (T) {\cal V}(T) =E$. This
implies, by (\ref{eq:inprep}), that
\andy{probinfu}
\beq
{\cal P}(T)=\mbox{Tr}\left[\rho_0{\cal V}^\dagger(T){\cal V}(T)\right]
= \mbox{Tr} \left[ \rho_0 E \right] = 1 .
\label{eq:probinfu}
\eeq
If the particle is ``continuously" observed,
in order to check whether it has survived inside ${\cal H}_E$ ,
it will never make a transition to  ${\cal H}-{\cal H}_E$.
This is the ``quantum Zeno paradox."

An important remark is now in order: the theorem just summarized
{\em does not} state that the system {\em remains}
in its initial state, after the series of very frequent measurements.
Rather, the system is left in the subspace ${\cal H}_E$,
instead of evolving ``naturally" in the total
Hilbert space ${\cal H}$. This subtle
point, implied by Eqs.\ (\ref{eq:infproie})-(\ref{eq:probinfu}),
is often not duely stressed in the literature.

Notice also the conceptual gap between
Eqs.\ (\ref{eq:probNob}) and (\ref{eq:probinfob}): To perform an
experiment with $N$ finite is only a practical problem, from the
physical point of view.
On the other hand, the $N \rightarrow \infty$ case
is physically unattainable, and is rather to be regarded as a
mathematical limit (although a very interesting one).
In this paper, we shall not be concerned with this problem
(thoroughly investigated in  \cite{NNPR}) and shall
consider the $N \to \infty$ limit
for simplicity. This will make the analysis more transparent.

\setcounter{equation}{0}
\section{Quantum Zeno effect with neutron spin }
\label{sec-neutr}
\andy{neutr}

The example we consider is a neutron spin in a magnetic
field \cite{PNBR}. (A photon analog was
first outlined by Peres \cite{Peres}.)
We shall consider two different experiments: Refer to Figures 1(a) and 1(b).
In the case schematized in Figure~1(a),
\begin{figure}
\begin{center}
\epsfig{file=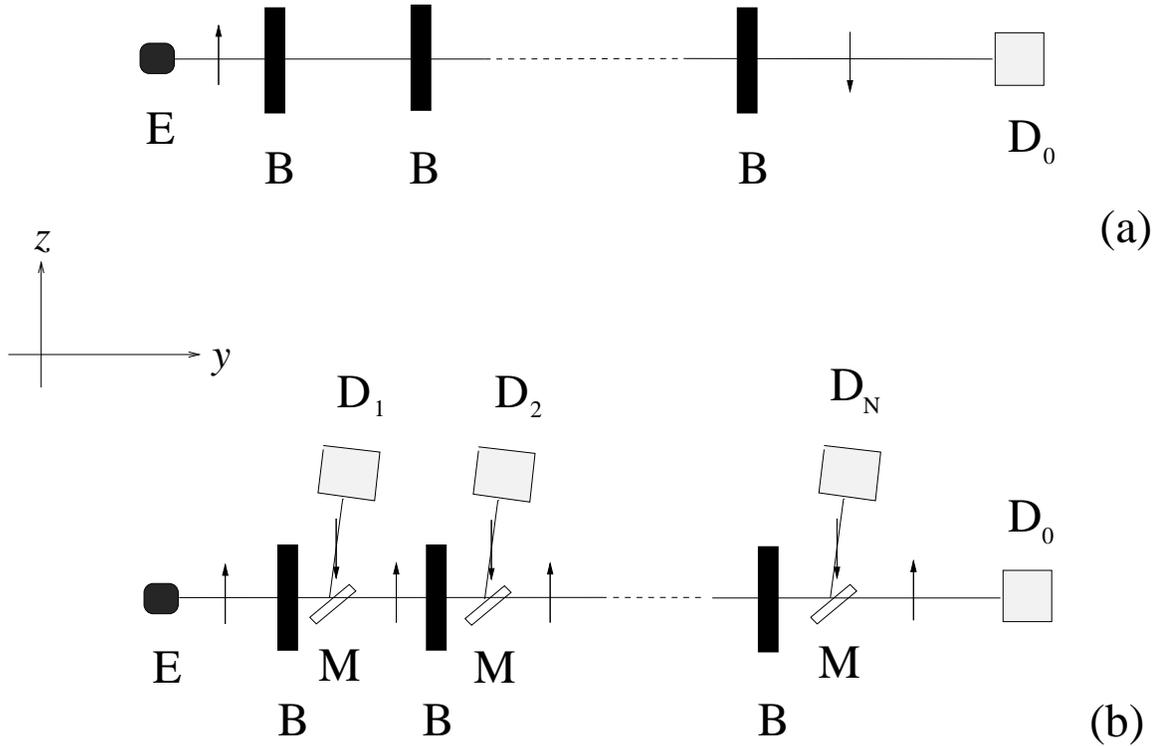,width=\textwidth}
\end{center}
\caption{(a) Evolution of the neutron spin
under the action of a magnetic field. An emitter sends a spin-up neutron
through several regions where a magnetic field $B$ is present.
The detector $D_0$ detects a spin-down neutron:
No Zeno effect occurs.
(b) Quantum Zeno effect: the neutron spin is ``monitored"
at every step, by selecting and detecting the spin-down component.
$D_0$ detects a spin-up neutron. }
\label{fig:fig1}
\end{figure}
the neutron interacts with several identical regions in which there is
a static magnetic field $B$, oriented along the $x$-direction.
We neglect here any losses and assume that
the interaction be given by the Hamiltonian
\andy{simpleH}
\beq
H= \mu B \sigma_1,
\label{eq:simpleH}
\eeq
$\mu$ being the (modulus of the) neutron magnetic moment,
and $\sigma_i \; (i=1,2,3)$ the Pauli matrices.
We denote the spin states of the neutron along the
$z$-axis  by $\vert \uparrow\rangle$ and $\vert \downarrow\rangle$. 

Let the initial neutron state be
$\rho_{0} = \rho_{\uparrow \uparrow} \equiv
\vert \uparrow \rangle\langle\uparrow \vert$.
The interaction with the magnetic field provokes a rotation of the
spin around the $x$-direction. After crossing the whole setup, the
final density matrix reads
\andy{finalstepT}
\beq
\rho (T) \equiv
e^{-iHT} \rho_{0} e^{iHT} =
\cos ^2 {\frac{\omega T}{2}} \rho\sb{\uparrow \uparrow}
 + \sin ^2{\frac{\omega T}{2}} \rho \sb{\downarrow \downarrow}
 - \frac i2\sin{\omega T}(
  \rho \sb{\uparrow \downarrow}- \rho \sb{\downarrow\uparrow }),
\label{eq:finalstepT}
\eeq
where $\omega=2 \mu B$ and $T$ is the total time spent in the $B$ field.
Notice that the free evolution is neglected (and so are reflection effects,
wave-packet spreading, etc.).
If $T$ is chosen so as to satisfy the ``matching" condition
$\cos \omega T/2 = 0$, we obtain
\andy{noZeno}
\beq
\rho (T) = \rho\sb{\downarrow \downarrow}
\qquad \quad \left( T = (2m+1) \frac{\pi}{\omega}, \;\; m \in {\bf N} \right),
\label{eq:noZeno}
\eeq
so that the probability
that the neutron spin is down at time $T$ is
\andy{pT}
\beq
P_{\downarrow}(T) =1
\qquad \quad \left( T = (2m+1) \frac{\pi}{\omega}, \;\; m \in {\bf N} \right).
\label{eq:pT}
\eeq
The above two equations correspond to Eqs.\ (\ref{eq:noproie}) and
(\ref{eq:stillun}).
In our example, $H$ is such that if the
system is initially prepared in the up state, it will evolve to
the down state after time $T$.
Notice that,
within our approximations, the experimental setup
described in Figure~1(a) is equivalent to the situation where a magnetic field
$B$ is contained in a single region of space.

Let us now modify the experiment just described by inserting at every step
a device able to select and detect one component [say the down
($\d$) one] of the neutron spin. This can be accomplished by
a magnetic mirror M and a detector D.
The former acts as a ^^ ^^ decomposer," by splitting a
neutron wave with indefinite spin (a superposed state of up
and down spins) into two branch waves
each of which is in a definite spin state
(up {\em or} down) along the $z$-axis. The down state is then forwarded to a
detector, as shown in Figure~1(b).
The magnetic mirror yields a spectral decomposition \cite{Wigner}
with respect to the spin
states, and can be compared to the inhomogeneous magnetic field in a
typical Stern-Gerlach experiment.

We choose the same initial state for Q as in the previous experiment
[Figure 1(a)]. The action of M+D is represented by
the operator $E \equiv \rho\sb{\uparrow \uparrow} $
[remember that we follow the evolution along the horizontal direction,
i.e.\ the direction the spin-up neutron travels,
in Figure~1(b)], so that
if the process is repeated $N$ times, like in Figure~1(b), we obtain
\andy{yesZeno}
\beq
\rho^{(N)}(T) = V_N(T) \rho_0 V_N^\dagger(T)
  = \left( \cos ^2 {\frac{\omega t}{2}} \right)^N
      \rho\sb{\uparrow \uparrow}
  = \left( \cos ^2 {\frac{\pi}{2N}} \right)^N
      \rho\sb{\u \u} ,
\label{eq:yesZeno}
\eeq
where the ``matching" condition for $T=Nt$ [see Eq.\
(\ref{eq:noZeno})] has been required again. The probability that
the neutron spin is up at time $T$, if $N$ observations have been
made at time intervals $t \; (Nt=T)$, is
\andy{pZT}
\beq
P_{\uparrow}^{(N)}(T) =
      \left( \cos ^2 {\frac{\pi}{2N}} \right)^N .
\label{eq:pZT}
\eeq

This discloses the occurrence of a QZE:
Indeed, $P_\uparrow^{(N)}(T)> P_\uparrow^{(N-1)}(T)$ for $N\geq 2$,
so that the evolution is ``slowed down" as $N$ increases. Moreover,
in the limit of infinitely many observations
\andy{NZeno}
\beq
\rho^{(N)}(T) \stackrel{N \rightarrow \infty}{\longrightarrow}
\t{\rho}(T) =  \rho\sb{\u \u}
\label{eq:NZeno}
\eeq
and
\andy{probfr}
\beq
{\cal P}_\u (T) \equiv \lim_{N \rightarrow \infty}
P_\u^{(N)}(T) = 1.
\label{eq:probfr}
\eeq
Frequent observations ``freeze" the neutron spin in its initial state, by
inhibiting ($N \geq 2$) and eventually hindering ($N \rightarrow \infty$)
transitions to other states.
Notice the difference from Eqs.~(\ref{eq:noZeno}) and (\ref{eq:pT}):
The situation is completely reversed.

\setcounter{equation}{0}
\section{The spatial degrees of freedom}
\label{sec-neutrspin}
\andy{neutrspin}

In the analysis of the previous section only the spin degrees
of freedom were taken into account.
No losses were considered, even though
their importance was already mentioned in \cite{PNBR,NNPR}.
In spite of such a simplification, the model yields
physical insight into the Zeno phenomenon, and has the nice
advantage of being solvable.

We shall now consider a more detailed description. The practical
realizability of this experiment has already been discussed, with
particular attention to the $N \rightarrow \infty$ limit and
various possible losses \cite{NNPR}. One source of losses is the
occurrence of reflections at the boundaries of the interaction
region and/or at the spectral decomposition step. A careful
estimate of such effects would require a dynamical analysis of the
motion of the neutron wave packet as it crosses the whole
interaction region (magnetic-field regions followed by field-free
regions containing each a magnetic mirror M that performs the
``measurement"). However, it is not an easy task to include the
spatial degrees of freedom of the neutron in the analysis; instead,
we shall adopt a simplified description of the system, which
preserves most of the essential features and for which an explicit
solution can still be obtained. It turns out that the inclusion of
the spatial degrees of freedom in the evolution of the spin state
can result in completely different situations from the ideal case,
which in turn clarifies the importance of losses in actual
experiments and, at the same time, sheds new light on the Zeno
phenomenon itself.

Let us now try to incorporate the other degrees of
freedom of the neutron state in our description.
Let our state space be the 4-dimensional Hilbert space
$\cH_p \otimes \cH_s$, where
$\cH_p = \{ |R \rangle, |L \rangle \}$ and
$\cH_s = \{ |\uu \rangle, |\dd \rangle \}$
are 2-dimensional Hilbert spaces, with $R (L)$ representing a particle
traveling towards the right (left) direction along the $y$-axis,
and $\uu (\dd)$ representing spin up (down) along the $z$-axis.
We shall set, in the respective Hilbert spaces,
\andy{settings}
\beq
  |R \rangle = \upp, \quad |L \rangle = \downn; \quad
  |\uu \rangle = \upp, \quad |\dd \rangle = \downn,
\label{eq:settings}
\eeq
so that, for example, the state $|R \dd \rangle$
 represents a spin-down particle traveling towards the right
direction ($+y$).
Also, for the sake of simplicity, we shall work with vectors, rather than
density matrices (the extension is straightforward).

In this extended Hilbert space the first
Pauli matrix $\sigma_1$ acts only on $\cH_s$ as a spin flipper,
$\sigma_1 |\uu \rangle = |\dd \rangle$ and
$\sigma_1 |\dd \rangle = |\uu \rangle$, while 
another first Pauli matrix $\tau_1$ acts only on $\cH_p$ as a
direction-reversal operator, 
$\tau_1|R\rangle=|L\rangle$ and $\tau_1|L\rangle=|R\rangle$.
To investigate the effects of reflection
we assume that the interaction be described by the Hamiltonian
\andy{modelH}
\beq
H= g (1 + \alpha \tau_1)(1 + \beta \sigma_1),
\label{eq:modelH}
\eeq
where $g, \alpha$ and $\beta$ are real constants.
By varying these parameters and the total
interaction time $T$, the above Hamiltonian can describe various
situations in which a neutron, impinging on a $B$-field applied
along $x$-axis, undergoes transmission/reflection and/or spin-flip
effects.

It is worth pointing out that the above Hamiltonian
incorporates the spatial degrees of freedom in an abstract way:
Only the 1-dimensional motion of the neutron, represented by $L$ and $R$,
has been taken into account and all other effects (like
for instance the spread of the wave packet) are neglected.
This amounts to consider a trivial free Hamiltonian, which can be
dropped out from the outset.
This may seem too drastic an approximation;
however, it is not as rough as one may imagine. In fact,
over the distances involved (a neutron interferometer), the
spread of the wave packet can always be practically neglected as a first
approximation.
The introduction of the above two degrees of freedom $L$ and $R$
just corresponds to such a situation and the simplicity
of the model still enables us to obtain explicit solutions for the
dynamical evolution. This can be a great advantage.
A realization of such a quantum Zeno effect experiment is in progress 
at the pulsed ISIS neutron spallation source. 
Neutrons which are trapped between perfect crystal plates pass on each 
of their 2000 trajectories through a flipper device which cause an adjustable 
spin rotation. 
Flipped neutrons immediately leave the storage system where they can be 
easily detected (see e.g.\ \cite{Jerica}).

\begin{figure}
\begin{center}
\epsfig{file=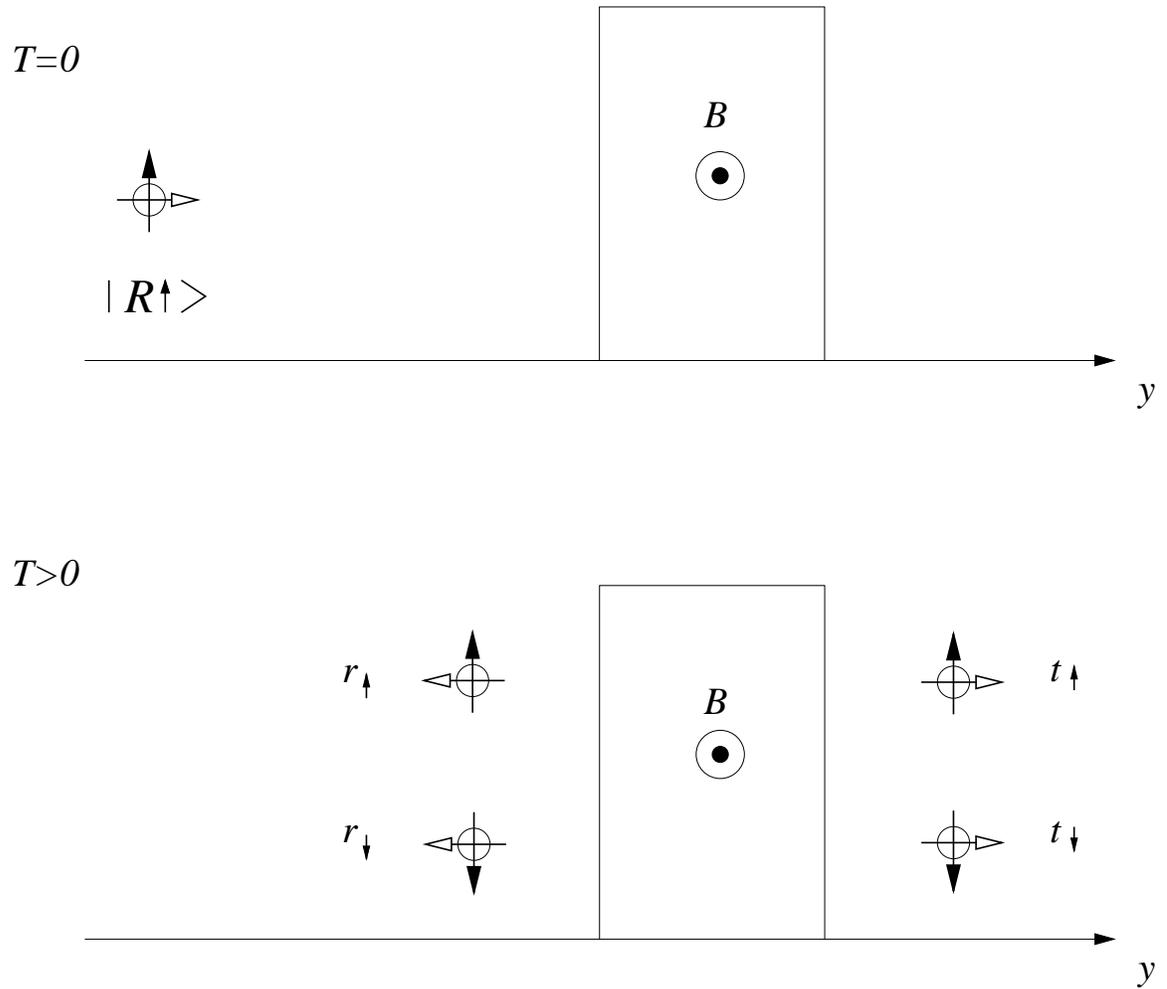,width=\textwidth}
\end{center}
\caption{Transmission and reflection coefficients for a neutron
initially prepared in the $|R \uu \rangle$ state. }
\label{fig:ssprob}
\end{figure}

Since the spin flipper $\sigma_1$ and the direction-reversal
operator $\tau_1$ commute with each other and with the Hamiltonian
(\ref{eq:modelH}), the energy levels of the system governed by this
Hamiltonian are obviously $E_{\tau\sigma} \equiv g
(1+\tau\alpha)(1+\sigma\beta)$ with $\tau,\sigma=\pm$. Moreover,
the evolution of the system has the following factorized structure
\begin{equation}\label{fs}
e^{-iHT}=e^{-igT}e^{-i\alpha gT\tau_1}e^{-i\beta gT\sigma_1}
e^{-i\alpha\beta gT\tau_1\sigma_1}.
\end{equation}
If a neutron is initially prepared in state $|R \uu \rangle$,
the evolution operator is explicitly expressed as
\andy{e-iHT}
\beq
e^{-iHT}=t_\uparrow
         +r_\uparrow\tau_1
         +t_\downarrow\sigma_1
         +r_\downarrow\tau_1\sigma_1,
\label{eq:e-iHT}
\eeq
where $t_\uu, t_\dd, r_\uparrow$ and $r_\downarrow$ are the
transmission/reflection coefficients of a neutron, whose spin is
flipped/not flipped after interacting with a constant magnetic
field $B$, applied along the $x$-direction in a finite region of
space (square potential, stationary state problem). See Figure
\ref{fig:ssprob}.
These coefficients are connected with the energy levels by the
following relation
\andy{corrs}:
\beq
\pmatrix{t_\uu & t_\dd \cr r_\uparrow & r_\downarrow}=\frac 14
\pmatrix{1 & 1 \cr 1 & -1}
\pmatrix{e^{-iE_{++}T} & e^{-iE_{+-}T} \cr e^{-iE_{-+}T} & e^{-iE_{--}T}}
\pmatrix{1 & 1 \cr 1 & -1}.
\label{eq:corrs}
\eeq

By specifying the values of the parameters $g, \alpha, \beta$ and
the total interaction time $T$, one univocally determines
$t_\uu, t_\dd, r_\uparrow$ and $r_\downarrow$.
Direct physical meaning can therefore be attributed to the constants
$g, \alpha$ and $\beta$ in (\ref{eq:modelH}) by comparison with the
transmission/reflection coefficients. For example, in order to mimic
a realistic experimental setup with given values of
$t_{\uu\dd}, r_{\uparrow\downarrow}$, it is enough to obtain the
values of $g, \alpha$ and $\beta$ from (\ref{eq:corrs}) and plug them
in the Hamiltonian (\ref{eq:modelH}).
The model could in principle be further improved by making the constant $g$
energy-dependent. We will consider a more realistic
Hamiltonian in Sec.~\ref{sec-numan}.

\setcounter{equation}{0}
\section{The ideal case of complete transmission}
\label{sec-compltrans}
\andy{compltrans}

In the following discussions we always assume that our initial
state is $|R\uu \rangle $, i.e., a right-going spin-up neutron, and
consider, for definiteness, the case of total transmission with
spin flipped, i.e., $|t_\dd|^2=1$, {\em when no measurements are
performed}. Of course, this has to be considered as an idealized
situation, since a spin rotation can only take place when there is
an interaction potential (proportional to the intensity of the
magnetic field) which necessarily produces reflection effects (with
the only exception of plane waves). Stated differently, when the
spatial degrees of freedom are taken into account in the scattering
problem off a spin-flipping potential, complete transmission is
impossible to achieve: There are always reflected waves. Our model
Hamiltonian (\ref{eq:modelH}) must therefore be regarded as a
simple caricature of the physical system we are analyzing. Wave
packet effects will be discussed in Sec.\ \ref{sec-numan}.

To obtain a total transmission with spin flipped, the evolution
operator should have the form $e^{-iHT}\propto\sigma_1$, which is
equivalent to either
\andy{either.or}
\begin{eqnarray}
&e^{-i\alpha gT\tau_1}\propto\tau_1,\quad e^{-i\beta gT\sigma_1}
\propto 1,\quad e^{-i\alpha\beta gT\tau_1\sigma_1}\propto
\tau_1\sigma_1,&
\label{eq:either} \\
\noalign{\noindent or}
&e^{-i\alpha gT\tau_1}\propto 1,\quad e^{-i\beta gT\sigma_1}
\propto \sigma_1,\quad e^{-i\alpha\beta gT\tau_1\sigma_1}\propto 1.&
\label{eq:or}
\end{eqnarray}
That is,
\andy{condi1,2}
\begin{eqnarray}
{\rm Case\ i)}\hfill&\cos\alpha gT=\sin\beta gT=\cos\alpha\beta gT=0,&\hfill
\label{eq:condi1}
\\
\noalign{\noindent or}
{\rm Case\ ii)}\hfill&\sin\alpha gT=\cos\beta gT=\sin\alpha\beta gT=0.&\hfill
\label{eq:condi2}
\end{eqnarray}
(All other cases, such as total reflection with/without spin-flip
can be analyzed in a similar way.)
In both cases, 
the evolution is readily computed:
\andy{evol0}
\beq
e^{-iHT} |R \uu \rangle  = \mbox{phase factor} \times |R \dd \rangle .
\label{eq:evol0}
\eeq
The boundary conditions are such that the neutron is
transmitted and its spin flipped with unit
probability. For the experimental realization, see \cite{spinflip}.
This is the situation outlined in Figure \ref{fig:fig1}(a).

We shall now focus on some interesting 
cases, which illustrate some definite aspects of the QZE.\ \
Let us see, in particular, how the evolution of the quantum state of the
neutron is modified by choosing different projectors (corresponding to
different ``measurements").

\subsection{Direction-insensitive spin measurement}
\label{sec-case1}
\andy{case1}

We perform now a series of measurements, in order to check whether the
neutron spin is up.
Let us call this type of measurement a ``direction-insensitive spin
measurement," for reasons that will become clear later.
\begin{figure}
\begin{center}
\epsfig{file=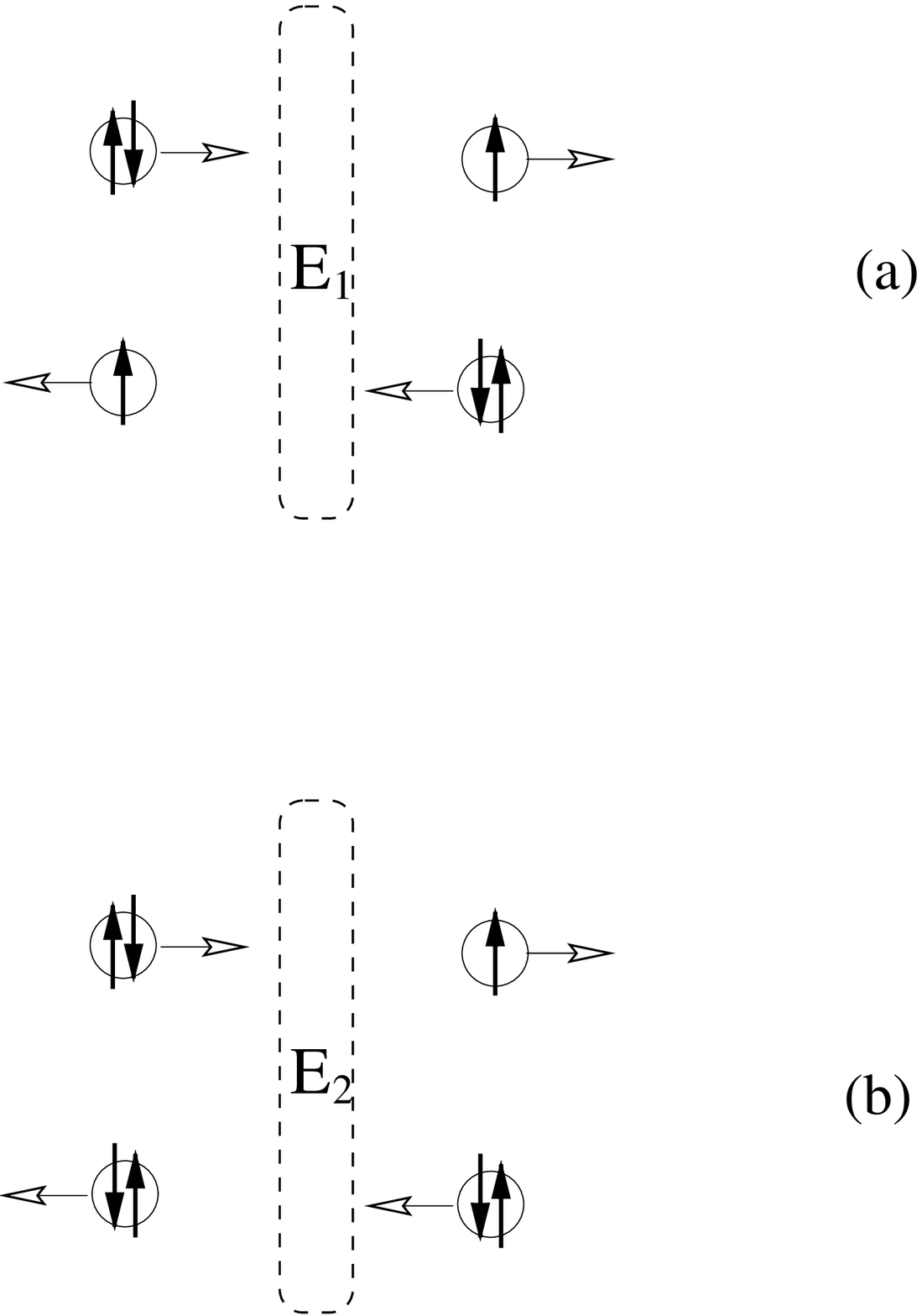,height=10cm}
\end{center}
\caption{(a) Direction-insensitive spin measurement.
(b) Direction-sensitive spin measurement.}
\label{fig:figins}
\end{figure}
Refer to Figure \ref{fig:figins}(a).
The projection operator corresponding to this measurement is
\andy{E1}
\beq
E_1=1-|R\d\rangle\langle R\d|-|L\d\rangle\langle L\d|={1\over2}(1+\sigma_3),
\label{eq:E1}
\eeq
that is, the spin-down components are projected out regardless of
the direction of propagation of the neutron. In this case, after
frequent measurements $E_1$ performed at time intervals $T/N$, the
evolution operator in Eq.~(\ref{eq:Nproie}) reads
\andy{VN1a}
\beq
V_{N}(T)
=\left( E_1e^{-iHT/N}E_1 \right)^N
=E_1(t_\uu+r_\uu\tau_1)^N ,
\label{eq:VN1a}
\eeq
where $t_\uu\sim1-igT/N$ and $r_\uu\sim-i\alpha gT/N$ for large $N$
[see Eq.\ (\ref{eq:e-iHT})]. Taking the limit, one obtains the
following expression for the QZE evolution operator defined in Eq.\
(\ref{eq:slim}):
\andy{V1(T)a}
\beq
{\cal V}(T)
=\lim_{N\to\infty}V_N(T)
 =e^{-igT}E_1e^{-i\alpha gT\tau_1}.
\label{eq:V1(T)a}
\eeq

Interesting physical situations can now be investigated. Choose,
for instance, $gT= \pi, \alpha= -1/2, \beta= -1$, which belongs to
Case i) in Eq.\ (\ref{eq:condi1}) [so that, without measurements,
the neutron is totally transmitted with its spin flipped, as shown
in Eq.\ (\ref{eq:evol0})]. When the direction-insensitive
measurements are continuously performed, the QZE evolution is
${\cal V}(T)=-iE_1\tau_1$ and the final state is
\andy{fst0}
\beq
{\cal V}(T)|R\uu\rangle=-i|L\uu\rangle,
\label{eq:fst0}
\eeq
i.e., {\em the neutron spin is not flipped, but the neutron itself
is totally reflected}. This clearly shows that reflection
``losses" can be very important; as a matter of fact, reflection effects
{\em dominate}, in this example. Notice that this is always an example
of QZE: The projection operator $E_1$ in
(\ref{eq:E1}) {\em prevents} the spin from flipping.
The point here is, however, that $E_1$ is
not ``tailored" so as to prevent the wave function from being reflected!

\subsection{Another particular case: seminal model}
\label{sec-semmod}
\andy{semmod}

Let us now focus on a model corresponding to
Case ii) in Eq.~(\ref{eq:condi2}). The choice of parameters, e.g.\
$gT=\pi/2, \alpha=2n, \beta=-1$,
obviously fulfills these conditions for
arbitrary integer $n$.
Total transmission with spin flipped occurs again when
no measurement is performed.

When direction-insensitive spin measurements, described by
projections $E_1$, are performed at time intervals $T/N$ and in the
$N\to \infty$ limit, the QZE evolution operator in
Eq.~(\ref{eq:V1(T)a}) becomes simply ${\cal V}_1(T)=-i(-1)^nE_1$
and the final state is
\beq\label{eq:fst00a}
{\cal V}_1(T)|R\u\rangle=-i(-1)^n|R\u\rangle,
\eeq
so that the ``usual" QZE is obtained. When $n=0$ this is our
seminal model \cite{PNBR}, reviewed in Sec.~\ref{sec-neutr}.
Obviously, the case $n=0$ is not rich enough to yield information
about reflection effects. In the following subsection the case of
nonzero $n$ will be discussed.

\subsection{Direction-sensitive spin measurements}
\label{sec-case2}
\andy{case2}

We now consider a different type of spin measurement.
Let the measurement be characterized by the
following projection operator
\andy{E2}
\beq
E_2=1-|R\dd\rangle\langle R\dd|,
\label{eq:E2}
\eeq
which projects out those neutrons that are transmitted with their
spin flipped. Notice that spin-down neutrons that are reflected are
not projected out by $E_2$: for this reason we call this a
``direction-sensitive" spin measurement. Refer to
Figure~\ref{fig:figins}(b). Even though the action of this
projection is not easy to implement experimentally, this example
clearly illustrates some interesting issues related to the
Misra--Sudarshan theorem. We shall see that the action of the
projector $E_2$ will yield a very interesting result. For large
$N$, the evolution is given by
\beq
V_{2,N}(T)=
\left(E_2e^{-iHT/N}E_2\right)^N =
e^{-igT}\left(1-i\frac{gT}{N} Z\right)^NE_2+O(1/N),
\eeq
where $Z\equiv E_2(H/g-1)E_2$.
The QZE evolution is given by the limit
\beq\label{eq:ev2}
{\cal V}_2(T)=\lim_{N\to\infty}V_{2,N}(T)=e^{-igT}e^{-igTZ}E_2 .
\eeq
To compute its effect on the initial state $|R\uparrow\rangle$,
we note that, when acting on states $|R\uparrow\rangle, |L\uparrow\rangle$
and $|L\downarrow\rangle$, which span the ``survival" subspace,
the $Z$ operator behaves as
\beq
Z \left(\matrix{|R\uparrow\rangle\cr |L\uparrow\rangle\cr
|L\downarrow\rangle}\right)
=\left(\matrix{0&\alpha&\alpha\beta\cr \alpha&0&\beta\cr
\alpha\beta&\beta&0}\right)
\left(\matrix{|R\uparrow\rangle\cr |L\uparrow\rangle\cr
|L\downarrow\rangle}\right).
\eeq
Let us choose for definiteness $\beta=-1$, so that
\beq
(Z-1/2)^2|R\u\rangle=\theta^2|R\u\rangle,
\label{eq:Z2Rup}
\eeq
with $\theta=\sqrt{8\alpha^{2}+1}/2$.
Thus the final state can be readily obtained
\andy{VTRup2}
\beq
{\cal V}_2(T)|R\u\rangle
=
e^{-3igT/2}\Biggl[
\left(\cos gT\theta+\frac{i}{2\theta}\sin gT\theta
 \right)|R\u\rangle
+\frac{i\alpha}{\theta}\sin gT\theta
\Bigl(|L\d\rangle-|L\u\rangle\Bigr)\Biggr].
\label{eq:VTRup2}
\eeq
Therefore, for a continuous direction-sensitive
(namely, $E_2$) measurement, the probability of
finding the initial state $|R\u\rangle$
is not unity. Part of the wave function will be
reflected, although the neutron would have been totally transmitted
without measurement [see (\ref{eq:evol0})] or with an ``$E_1$-measurement"
[see (\ref{eq:fst00a})].

Clearly, the action of the projector $E_2$ yields a completely
different result from that of $E_1$, in (\ref{eq:fst00a}). This is
obvious and easy to understand: the state (\ref{eq:VTRup2}) belongs
to the subspace of the ``survived" states, {\em according to the
projection $E_2$.} Notice also that the probability loss due the
measurements is zero, in the limit, because the QZE evolution
(\ref{eq:ev2}) is unitary within the subspace of the \lq\lq
survived" states.

\setcounter{equation}{0}
\section{A more realistic model }
\label{sec-numan}
\andy{numan}

Let us now introduce a more realistic (albeit static) model. Such a
model can be shown to be derivable from a Hamiltonian very similar
to the one studied in the previous sections by a suitable
identification of parameters (see Appendix A). The effect of
reflections in the QZE will now be tackled by directly solving a
stationary Schr\"odinger equation, which will be set up as follows.

\begin{figure}
\epsfig{file=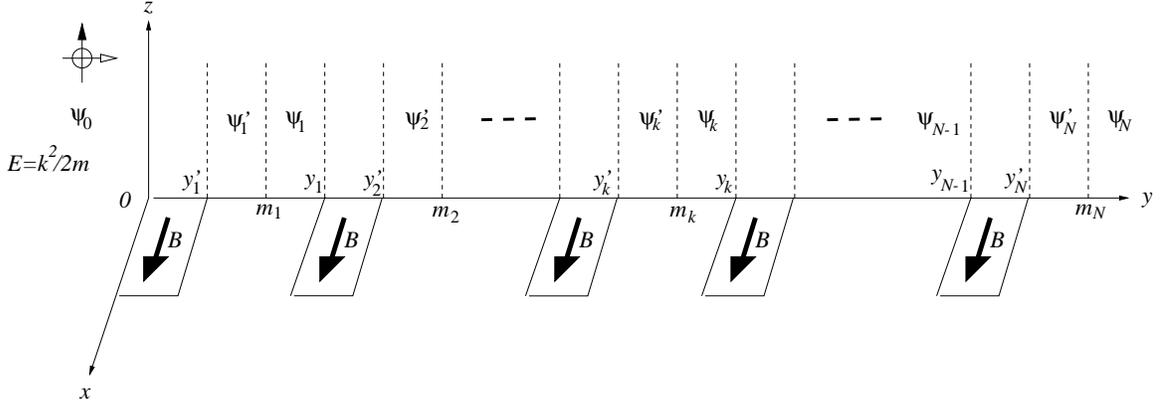,width=\textwidth}
\caption{Spin-up neutron moving along the $+y$ direction with energy
$E$. The magnetic field points to the $+x$ direction and is zero in
the region between $y_n^\prime$ and $y_n$, in which the
measurements will be made. In these field-free regions the wave
functions are $|\psi_n^\prime\rangle$ before measurement and
$|\psi_n\rangle$ after the measurement.}
\label{fig:figYu1}
\end{figure}

Let a neutron with energy $E=k^2/2m$ and spin up ($+z$ direction),
moving along the $+y$ direction, impinge on $N$ regions of constant
magnetic field pointing to the $x$ direction, among which there are
$N-1$ field-free regions. The thickness of a single piece of
magnetic field is $a$ and the field-free region has size $b$. The
configuration is shown in Figure \ref{fig:figYu1}.
Thus we have the one-dimensional scattering problem of a neutron
off a piecewise constant magnetic field with total thickness
$D=Na$. The stationary Schr\"odinger equation is described by the
Hamiltonian
\andy{realH}
\beq
H_{\rm Z}=\frac{p_y^2}{2m}+\mu B\sigma_1\Omega(y),
\label{eq:realH}
\eeq
where $\mu$ is the modulus of the neutron magnetic momentum, $B$
the strength of the magnetic field and
\beq
\Omega(y)
=\left\{\matrix{0&\mbox{for}&y<0,&y_n^\prime<y<y_n,   &y_N^\prime<y&
                (n=1,2,\ldots,N),\cr
\noalign{\smallskip}
                1&\mbox{for}&    &y_{n-1}<y<y_n^\prime&            &
                (n=1,2,\ldots,N),\cr}\right.
\eeq
with $y_n=n(a+b)$ and $ y_{n}^\prime=y_{n-1}+b$, characterizes the
configuration of the magnetic field $B$ applied along the $x$-axis.
Refer to Figure~\ref{fig:figYu1}. The incident state of the neutron
is taken to be $|\psi_{\rm in}\rangle=e^{iky}|\u\rangle$.
Let $r_{\u(\d)}$ be the reflection amplitude for the spin-up (spin-down)
component. The wave function for $y<0$ is written as
\beq\label{w1}
|\psi_0\rangle=e^{iky}|\u\rangle+e^{-iky}[r_{\u}|\u\rangle+r_\d|\d\rangle].
\eeq
Denoting the transmission amplitudes for spin-up and spin-down as
$t_\u$ and $t_\d$, the outgoing wave function in the region
$y>y_N^\prime$ reads
\beq\label{w2}
|\psi_N\rangle= e^{iky}[t_\u|\u\rangle+t_\d |\d\rangle].
\eeq
Since $[\sigma_1,H_{\rm Z}]=0$, it is convenient to work with the
basis $|\pm\rangle=(|\u\rangle\pm|\d\rangle)/\sqrt 2$, 
i.e., the eigenstates of $\sigma_1$ belonging to eigenvalues $\pm 1$.
For later use we denote $r_\pm=r_\u\pm r_\d$ and
$t_\pm=t_\u\pm t_\d$.

In the field-free region, before the point $y=m_n$ where the $n$th
measurement is assumed to take place, $y_n^\prime<y<m_n$, the wave
function is
\beq
|\psi_n^\prime\rangle= \sum_{\sigma=\pm}
\left(R^\prime_{n,\sigma}e^{ik(y-y^\prime_n)}+
L^\prime_{n,\sigma}e^{-ik(y-y^\prime_n)}\right)|\sigma\rangle,\quad
(n=1,2,\ldots,N).
\eeq
On the other hand, in the region after the $n$th measurement,
$m_n<y<y_n$, the wave function is
\beq
|\psi_n\rangle= \sum_{\sigma=\pm}\left(R_{n,\sigma}e^{ik(y-y_n)}+
                       L_{n,\sigma}e^{-ik(y-y_n)}\right)
                       |\sigma\rangle,\quad
(n=0,1,\ldots,N).
\eeq

The relation between the amplitudes of the wave functions
$|\psi^\prime_{n+1}\rangle$ and $|\psi_n\rangle$ at the right- and
left-hand sides of the $n$th potential region is determined by the
boundary conditions at points $y_n$ and $y_n^\prime$. In fact, we
have
\andy{ev1}
\beq
\label{ev1}
\pmatrix{R^\prime_{n+1,\pm}\cr L^\prime_{n+1,\pm}}=
M_\pm\pmatrix{R_{n,\pm}\cr L_{n,\pm}},
\eeq
where the transfer matrix is given by
\andy{Mpm}
\beq\label{eq:Mpm}
M_\pm =\pmatrix{
\cos k_\pm a+i\cosh\eta_\pm\sin k_\pm a
&-i\sinh\eta_\pm\sin k_\pm a\cr
i\sinh\eta_\pm\sin k_\pm a &
\cos k_\pm a-i\cosh\eta_\pm\sin k_\pm a }
\eeq
with $k_\pm=\sqrt{k^2\mp 2m\mu B}$ and $k/k_\pm=e^{\eta_\pm}$. This
can also be expressed in a concise way in terms of the Pauli
matrices in this two dimensional space, $\tau_1$, $\tau_2$ and
$\tau_3$, as
\beq
M_\pm= e^{\eta_\pm(1+\tau_1)/2}e^{ik_\pm a\tau_3}e^{-\eta_\pm(1+\tau_1)/2}.
\eeq
We clearly see that the above formula 
contains all the boundary information at
points $y_n$ and $y_n^\prime$: The first and last factors are the
kicks exerted at the boundaries of a single piece of constant
magnetic field, while the central one represents a free evolution
with relative energy $E\mp\mu B$.

In what follows, we shall incorporate the measurement processes
performed at points $m_n$ as some kind of boundary conditions,
connecting the primed and unprimed wave functions in the field-free
region.

\subsection{Evolution without any spin measurements}
\label{sec-nomeas}
\andy{nomeas}

We first consider the case where there is no measurement at all.
This enables us to set up the notation and rederive some known
results (which will be useful for future comparison). In this case
the primed and unprimed wave functions must be equal
$|\psi^\prime_n\rangle=|\psi_n\rangle$ in the field-free region. By
virtue of Eq.~(\ref{ev1}) we obtain
\beq\label{eq:rel1}
\pmatrix{R_{n+1,\pm}\cr L_{n+1,\pm}}=e^{ikb\tau_3}M_\pm
\pmatrix{R_{n,\pm}\cr L_{n,\pm}}.
\eeq
Notice the boundary
conditions $R_{0,\pm}=1/\sqrt 2$ and 
$L_{N,\pm}=0$ together with the definitions of transmission amplitude
$R_{N,\pm}=e^{iky_N}t_{\pm}/\sqrt 2$ and reflection amplitude
$L_{0,\pm}=r_\pm/\sqrt 2$.
After applying the above equation $N$ times, we obtain the
following relation
\beq\label{eq:trans}
e^{iky_N}\pmatrix{t_{\pm}\cr 0\cr}
=\left([N]_\pm e^{ikb\tau_3}M_\pm-[N-1]_\pm\right)
\pmatrix{1\cr r_{\pm}\cr},
\eeq
where $[N]_\pm=(q_\pm^N-q_\pm^{-N})/(q_\pm-q_\pm^{-1})$, with
$q_\pm, q_\pm^{-1}$ being the two eigenvalues of the transfer
matrix $e^{ikb\tau_3}M_\pm$, which are determined by
\beq\label{eq:trace}
\frac{q_\pm+q_\pm^{-1}}2=\cos kb\cos k_\pm a-\cosh\eta_\pm\sin kb\sin k_\pm a.
\eeq

When there is only a single piece of magnetic
field with length $a$, i.e., $N=1$, the transmission amplitude of
a neutron in the spin state $|\pm\rangle$ is
\beq
t_{a\pm}=\frac {e^{-ika}}
{\cos k_\pm a-i\cosh\eta_\pm\sin k_\pm a},
\eeq
as is well known. From Eq.\ (\ref{eq:trans}), for an arbitrary
$N>1$, the transmission amplitude of the same neutron passing
through a magnetic field with a lattice-like structure as depicted
in Figure 4 can be written as
\beq
t_\pm
=\frac {e^{-iky_N}t_{a\pm}}{e^{-iky_1}[N]_\pm -[N-1]_\pm t_{a\pm}}.
\eeq
For a neutron in its spin-up state $|\u\rangle$, the transmission
amplitude with spin unflipped is then $t_\u=(t_++t_-)/2$ and that
with spin flipped is $t_\d=(t_+-t_-)/2$. As a result, for a spin-up
neutron to go through a constant potential of width $y_N=D=Na$
without reflection and with spin flipped, i.e., $|t_\d|=1$, one
should require $k_\pm D=n_\pm\pi$ or
\beq\label{ttc}
E=\frac{\pi^2(n_+^2+n_-^2)}{4mD^2},\quad
\mu B=\frac{\pi^2(n_+^2-n_-^2)}{4m D^2},
\eeq
with $n_\pm$ two arbitrary integers, their difference $n_+-n_-$ 
being an odd number. In this case of complete transmission
$|t_\d|=1$, the energy $E$ must be larger than the potential $\mu
B$. The rest of the analysis above, however, is valid also when the
energy is less than the potential.

Now we consider the case where $N$ tends to infinity and the
magnetic field possesses a periodic lattice structure. The relation
(\ref{eq:rel1}) still holds and in order to preserve the
translational symmetry along the $y$ axis (that is, to keep the 
Hamiltonian invariant under a translation of $(a+b)$ along the 
$y$-axis), one should have 
$|q_\pm|=1$ owing to the Bloch theorem. 
Equivalently, the trace of the transfer matrix $e^{ikb\tau_3}M_\pm$
as given in Eq.~(\ref{eq:trace}) should be less than one. This
determines the energy band of the system: those energies that make
the absolute value of this trace greater than 1 will be forbidden,
because for these energies $|q_\pm|$ or $|q_\pm|^{-1}$ becomes
larger than one and $[N]_\pm$ tends exponentially to infinity when
$N$ approaches infinity. For large $N$, even if there is no
periodical structure, there is always some $k$ that makes this
trace greater than one (e.g. $kb+k_\pm a=l\pi$). Therefore the
transmission probability will tend to zero exponentially when $N$
becomes large, even though the energy may be very large relative to
the potential. This shows that reflection effects in presence of a
lattice structure are very important; as we shall see, this feature
is preserved even when projection operators are interspersed in the
lattice.

\subsection{Direction-insensitive projections}
\label{sec-insensitive}
\andy{insensitive}

We consider now the second situation, when direction-insensitive
measurements are performed at points $m_n$s. By this kind of
measurement, the spin-down components are projected out and the
spin-up components evolve freely regardless whether the neutron is
travelling right or left.

The boundary conditions imposed by this kind of measurement at point
$m_n$ for the wave function $|\psi_n\rangle$ and $|\psi^\prime_n\rangle$
in the field-free region are expressed as
\beq
R_{n,\d}=L^\prime_{n,\d}=0,
\quad \pmatrix{R_{n,\u}^\prime \cr L_{n,\u}^\prime \cr}
=e^{-ikb\tau_3}\pmatrix{R_{n,\u}\cr L_{n,\u} \cr},
\eeq
where $R_{n,\u}=(R_{n,+}+R_{n,-})/2$ and
$R_{n,\d}=(R_{n,+}-R_{n,-})/2$ for right-going components and
similar expressions for the left-going and primed components.
Therefore, application of Eq.~(\ref{ev1}) $N$ times yields
\beq\label{ev2}
\pmatrix{R_{N,\u}\cr L_{N,\u} \cr}
=(e^{ikb\tau_3}M_1)^N\pmatrix{R_{0,\u}\cr L_{0,\u} \cr},
\eeq
where the $2\times 2$ transfer matrix $M_1$ has the
following matrix elements
\beq
(M_1)_{ij}
=\bar{M}_{ij}-\Delta M_{i2}\Delta M_{2j}/\bar{M}_{22}
\eeq
with $\bar M=(M_++M_-)/2$ and $\Delta M=(M_+-M_-)/2$.

Now we take the limit as required by a ``continuous" measurement,
i.e., $N\to\infty$, $a\to0$ keeping $Na=D$ finite and $Nb\to 0$. By
the definition (\ref{eq:Mpm}) of the transfer matrix, we have the
small-$a$ expansions
\beq
\bar M=1+ika\tau_3+O(a^2),\quad
\Delta M= \zeta ka(\tau_2-i\tau_3)+O(a^2)
\eeq
with $\zeta\equiv\mu B/2E$, obtaining
\beq
\lim_{N\to\infty}(e^{ikb\tau_3}M_1)^N= e^{ikD\tau_3}.
\eeq
Recall that $t_\u=e^{-ikD}R_{N,\u}$ is the transmission amplitude,
$L_{0,\d}=r_\d$ the reflection amplitude and $L_{N,\u}=0$ and
$R_{0,\u}=1$ because of the boundary conditions. After taking the
limit $N\to\infty$ in Eq.~(\ref{ev2}), we see that the transmission
(survival) probability becomes one, i.e., $|t_{\u}|^2=1$, for {\em
any} input energy and magnetic field. This reveals another aspect
of neutron QZE: When the energy of the neutron is smaller than the
potential, the transmission probability decays exponentially when
the length increases and no measurement is performed; By contrast,
when continuous direction-insensitive measurements are made, one
can obtain a total transmission!

If we choose the energy of the neutron and the potential as in
Eq.~(\ref{ttc}), without measurements the neutron will be totally
transmitted with its spin flipped. On the other hand, if the
spin-up state is measured continuously, the neutron will be totally
transmitted with its spin unflipped. This is exactly the QZE in the
usual sense. Our analysis enables us to see that two kinds of QZEs
are taking place: One is the QZE for the right-going neutron, by
which we obtain a total transmission of the right-going input
state, and another one is for the left-going neutron, which
preserves the zero amplitude of the left-going input state. This
case corresponds to projector $E_1$ in our simplified model in
Sec.~\ref{sec-semmod}.

\subsection{Direction-sensitive projections}
\label{sec-sensitive}
\andy{sensitive}

The third case we consider is the direction-sensitive measurement.
By this kind of measurement the left-going components (or the
reflection parts) evolve freely, no matter whether spin is up or
down, and the right-going components are projected to the spin-up
state. The corresponding boundary conditions are
\beq
R_{n,\d}=0,\qquad L_{n,\pm}=e^{-ikb}L^\prime_{n,\pm}.
\eeq

If we apply Eq.~(\ref{ev1}) $N$ times, supplemented with these
boundary conditions, the following relations among the transmission
and reflection amplitudes are obtained
\beq\label{ev3}
e^{ikD}\pmatrix{t_{\u}\cr 0\cr 0}=\left(e^{ikb\Sigma_3}M_2\right)^N
\pmatrix{1\cr r_{\u}\cr r_{\d}},
\eeq
where $\Sigma_3$ is a diagonal matrix $\Sigma_3={\rm diag}\{1,-1,-1\}$
and the $3\times3$ transfer matrix $M_2$ is given by
\beq
M_2=\pmatrix{\bar{M}_{11}&\bar{M}_{12}&\Delta M_{12}\cr
                            \bar{M}_{21}&\bar{M}_{22}&\Delta M_{22}\cr
                            \Delta M_{21}&\Delta M_{22}&\bar{M}_{22}\cr}.
\eeq

In the limit of continuous measurements ($N\to\infty$, $a\to 0$,
while keeping $D=Na$ constant, and $Nb\to 0$), the transfer matrix
is expanded as
\beq
M_2=1-ika/3+ika Z_2+O(a^2),
\eeq
for small $a$, with ($\zeta=\mu B/2E$)
\beq\label{z2}
Z_2\equiv\pmatrix{4/3&0&-\zeta\cr
     0&-2/3&\zeta\cr
     \zeta&\zeta&-2/3\cr},
\eeq
and we have
\beq
\lim_{N\to\infty}(e^{ikb\Sigma_3}M_2)^{N}=e^{-ikD/3}e^{ikD Z_2}.
\eeq

\begin{figure}
\begin{center}
\epsfig{file=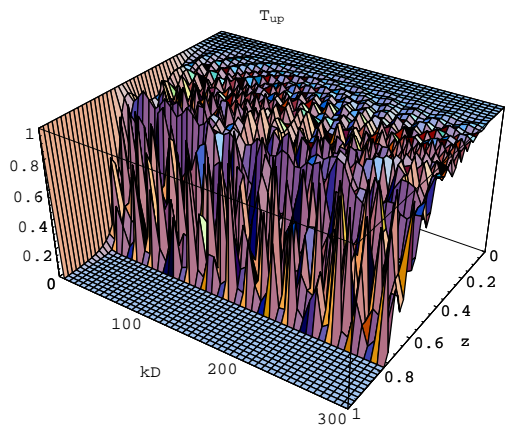,height=8cm}
\end{center}
\begin{center}
\epsfig{file=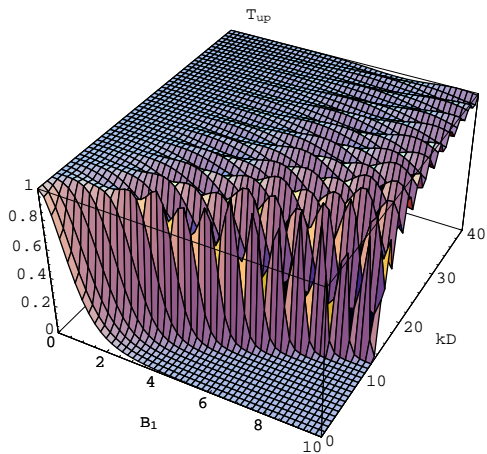,height=8cm}
\end{center}
\caption{The transmission probability with spin unflipped
$T_{\rm up}=|t_\u|^2$ is plotted as a function of $kD$ and $z=\zeta$ in
(a) and as a function of
$B_1=\sqrt{m\mu B}D$ 
and $kD$ in (b).}
\label{fig:figYu2}
\end{figure}

Notice that the matrix $Z_2$ satisfies $\Sigma_3
Z_2\Sigma_3=Z_2^\dagger$, from which we obtain, in the above limit,
the conservation of probability
\beq
|t_\u|^2+|r_\u|^2+|r_\d|^2=1.
\eeq
This shows that there are no losses caused by the continuous
direction-sensitive measurements. On the other hand, the
transmission amplitude with spin unflipped is explicitly given by
\beq
t_\u={{e^{-i4kD/3}}\over\left(e^{-ikD Z_2}\right)_{11}},
\eeq
which implies that the transmission probability $|t_\u|^2$ is in
general {\em not} equal to one. To have a general impression of its
behavior, we plot $T_{\u}=|t_\u|^2$ as a function of $kD$ and
$\zeta$ in Figure~\ref{fig:figYu2}.

Some comments are in order. There are two critical values for
$\zeta$, namely $0$ and $\zeta_c=4\sqrt3/9\approx 0.77$. When $0\le
\zeta<\zeta_c$, the matrix $Z_2$ has three real eigenvalues and the
transmission probability will oscillate depending on $kD$. When
$\zeta=\zeta_c$ the transmission probability will decay according
to $(kD)^{-2}$.
In fact, if one defines $G=Z_2-2/3$, it is easy to show that 
$e^{-ikDG}=1-ikDG+(e^{2ikD}-1-2ikD)G^2/4$, because $G$ satisfies 
$G^2(G+2)=0$.
Then one can explicitly confirm that the element $(e^{-ikDG})_{11}$ 
includes a linear $kD$ term, which gives the $(kD)^{-2}$ behavior to 
the transmission probability. Finally, when $\zeta>\zeta_c$ the
matrix $Z_2$ has two imaginary eigenvalues and therefore the
transmission probability decays exponentially with $kD$. This can
be seen clearly in Figure~5(a). An interesting case arises when we
consider $1/2<\zeta<\zeta_c$ or $E<\mu B<8\sqrt3E/9\approx 1.5E$.
Without measurements, the transmission probability decays
exponentially when the length of the magnetic field is increased,
because the input energy is smaller than the potential. When
continuous measurements are performed, however, the transmission
probability will oscillate as the length of the magnetic field
increases.

\begin{figure}
\begin{center}
\epsfig{file=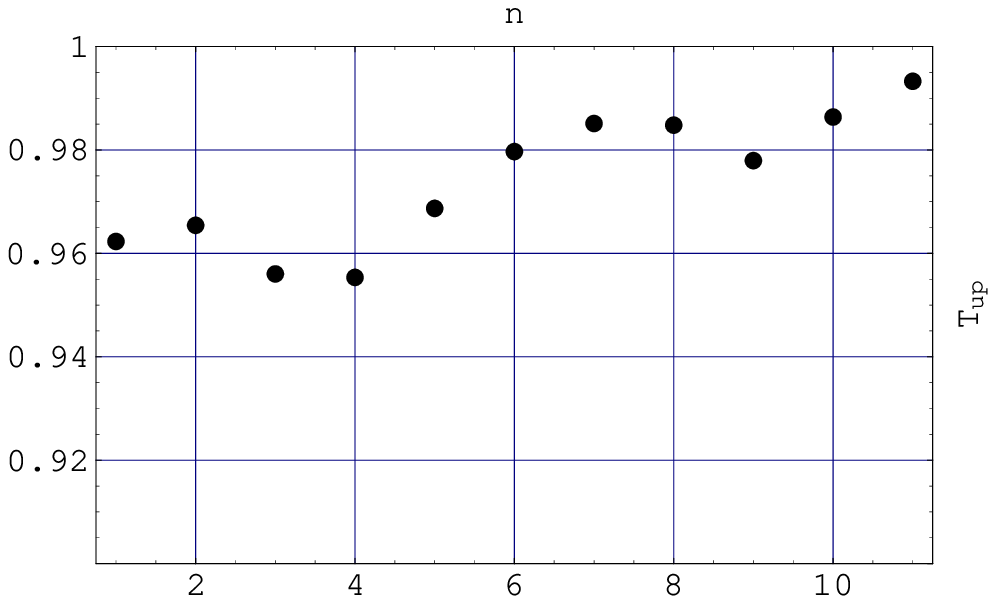,width=\textwidth}
\end{center}
\caption{The transmission probablity with spin unflipped $T_\uparrow=|t_\u|^2$
as a function of $n$, when the conditions 
 (\ref{ttc}) 
for total transmission are satisfied with $n_-=n$ and $n_+=n+9$.}
\label{fig:figYu3}
\end{figure}

As we can see in Figure~\ref{fig:figYu3}, although the conditions
(\ref{ttc}) for total transmission in absence of measurements have
been imposed, the transmission probability $T_{\u}$ is not one, as
it would be for the ``ordinary" QZE. Reflections are unavoidable.
This case corresponds to the projector $E_2$ considered in the
simplified model.

As we have seen, there are peculiar reflection effects in presence
of projections, when $D$ (total length) is varied. This is clearly
an interference effect, which can lead to enhancement of reflection
``losses," if the ``projection" does not suppress the left
component of the wave (this is what happens for $E_2$). This proves
that reflection effects can become very important in experimental
tests of the QZE with neutron spin, if, roughly speaking, the total
length of the interaction region ``resonates" with the neutron
wavelength. It is interesting that such a resonance effect takes
place even though the dynamical properties of the system are
profoundly modified by the projection operators, in the limit of
``continuous" measurements, leading to the QZE.

Finally, we would like to stress again that we are performing an
analysis in terms of stationary states (i.e.,
transmission/reflection coefficients for plane waves), while at the
same time we are analyzing a quantum Zeno phenomenon, which is
essentially a time-dependent effect. This is meaningful within our
approximations, where the wave-packet spread is neglected and the
measurements are performed with very high frequency. A more
sophisticated argument in support of this view is given in Appendix
A. In the present context wave packets effects, if taken into
account, would result in a sort of average of the effects shown in
Figures 5 and 6 (which refer to the monochromatic case); however,
our general conclusions would be unaltered. It is worth stressing
that, in neutron optics, effects due to a high sensitivity to
fluctuation phenomena (such as fluctuations of the intensity of the
magnetic field) become important at high wave number and constitute
an experimental challenge \cite{fluctB}.

\setcounter{equation}{0}
\section{Summary}
\label{sec-findisc}
\andy{findisc}

We have analyzed some peculiar features of a quantum Zeno-type
dynamics by discussing the noteworthy example of a neutron spin
evolving under the action of a magnetic filed in presence of
different types of measurements (``projections").

The ``survival probability" depends on our definition of
``surviving," i.e., on the choice of the projection operator $E$.
Different $E$s will yield different final states, and Misra and
Sudarshan's theorem \cite{Misra} simply makes sure that the
survival probability is unity: the final state belongs to the
subspace of the survived products.

In the physical case considered (neutron spin), our examples
clarify that the practical details of the experimental procedure by
which the neutron spin is ``measured" are very important. For
example, in order to avoid constructive interference effects,
leading to (unwanted) enhancement of the reflected neutron wave, it
is important to devise the experimental setup in such a way that
reflection effects are suppressed.

\section*{Acknowledgments}
This work is partly supported by the Grant-in-Aid for International 
Scientific Research: Joint Research \#10044096 from the Japanese 
Ministry of Education, Science and Culture, by Waseda University 
Grant for Special Research Projects No.~98A--619 and by the 
TMR-Network of the European Union ``Perfect Crystal Neutron Optics"
ERB-FMRX-CT96-0057.



\renewcommand{\thesection}{\Alph{section}}
\setcounter{section}{1}
\setcounter{equation}{0}
\section*{Appendix A}
\label{sec-appA}
\andy{appA}

\renewcommand{\thesection}{\Alph{section}}
\renewcommand{\thesubsection}{{\it\Alph{section}.\arabic{subsection}}}
\renewcommand{\theequation}{\thesection.\arabic{equation}}
\renewcommand{\thefigure}{\thesection.\arabic{figure}}

In this appendix, we shall endeavor to establish a connection
between the models analyzed in Secs.~\ref{sec-neutrspin} and
\ref{sec-numan}. In other words, we will examine whether the
parametrization of the Hamiltonian of the form (\ref{eq:modelH}) is
compatible with the more realistic one considered in
(\ref{eq:realH}) and in such a case find which values are to be
assigned to the parameters $\alpha,\beta$ and $g$. To this end, it
is enough to consider the scattering (i.e., the transmission and
reflection) process of a neutron off a single constant magnetic
field $B$ of width $a$. We compare the scattering amplitudes
calculated on the basis of the simple abstract Hamiltonian
(\ref{eq:modelH}) and of the more realistic one (\ref{eq:realH}).
Notice that the process is treated as a dynamical one in the former
case ($T$ is regarded, roughly speaking, as the time necessary for
the neutron to go through the potential), while in the latter case
we treat it as a stationary problem.

Observe first that the tranfer matrix $M_\pm$ in (\ref{eq:Mpm}),
derived for the stationary scattering process, yields the following
transmission/reflection amplitudes
\andy{RRLL}
\barr
R'_{1,\u}
={1\over2}\left({1\over(M_+)_{22}}+{1\over(M_-)_{22}}\right),&&
R'_{1,\d}
={1\over2}\left({1\over(M_+)_{22}}-{1\over(M_-)_{22}}\right),\nonumber\\
&&\label{eq:RRLL}\\
L_{0,\u}
=-{1\over2}\left({(M_+)_{21}\over(M_+)_{22}}
+{(M_-)_{21}\over(M_-)_{22}}\right),
&& L_{0,\d}
=-{1\over2}\left({(M_+)_{21}\over(M_+)_{22}}
-{(M_-)_{21}\over(M_-)_{22}}\right).\nonumber
\earr
It is easy to show that the relations (\ref{eq:RRLL}) are
equivalent to
\andy{corrs'}
\beq
\left(
\begin{array}{cccc}
1 & 1 & 1 & 1 \\
1 & -1 & 1 & -1 \\
1 & 1 & -1 & -1 \\
1 & -1 & -1 & 1
\end{array}
\right)
\left(
\begin{array}{c}
R'_{1,\u}\\
R'_{1,\d}\\
L_{0,\u}\\
L_{0,\d}
\end{array}
\right)
=
\left(
\begin{array}{c}
{\cal M}_{-,+}\\
{\cal M}_{-,-}\\
{\cal M}_{+,+}\\
{\cal M}_{+,-}
\end{array}
\right),
\label{eq:corrs'}
\eeq
where we have introduced
\andy{calMpm}
\beq
{\cal M}_{+,\pm}={1+(M_\pm)_{21}\over(M_\pm)_{22}},\quad
{\cal M}_{-,\pm}={1-(M_\pm)_{21}\over(M_\pm)_{22}}.
\label{eq:calMpm}
\eeq
It is important to realize that these quantities are just phase
factors. In fact, since
\andy{Melements}
\beq
(M_\pm)_{21}=i\sinh\eta_\pm\sin k_\pm a
\quad\mbox{and}\quad
(M_\pm)_{22}=\cos k_\pm a-i\cosh\eta_\pm\sin k_\pm a
\label{eq:Melements}
\eeq
and
\andy{abs2}
\beq
|1\pm(M_\pm)_{21}|^2=|(M_\pm)_{22}|^2=1+\sinh^2\eta_\pm\sin^2k_\pm a,
\label{eq:abs2}
\eeq
their absolute values are unity. Thus we can rewrite them in the
form
\andy{calMpp}
\beq
{\cal M}_{+,\pm}=e^{i(\xi_\pm+\phi_\pm)},\quad
{\cal M}_{-,\pm}=e^{i(-\xi_\pm+\phi_\pm)},
\label{eq:calMpp}
\eeq
where
\andy{phases}
\beq
\xi_\pm=\tan^{-1}(\sinh\eta_\pm\sin k_\pm a)\quad\mbox{and}\quad
\phi_\pm=\tan^{-1}(\cosh\eta_\pm\tan k_\pm a).
\label{eq:phases}
\eeq
Observe now that (\ref{eq:corrs}), dynamically derived from the
abstract Hamiltonian (\ref{eq:modelH}), is equivalent to
\andy{trvsgT}
\beq
\left(\matrix{t_\uu\cr
              t_\dd\cr
              r_\uu\cr
              r_\dd\cr}
\right)
={1\over4}\left(\matrix{1&1&1&1\cr
                         1&-1&1&-1\cr
                         1&1&-1&-1\cr
                         1&-1&-1&1\cr}
           \right)
 \left(\matrix{e^{-iE_{++}T}\cr
               e^{-iE_{+-}T}\cr
               e^{-iE_{-+}T}\cr
               e^{-iE_{--}T}\cr}
 \right).
\label{eq:trvsgT}
\eeq
The apparent similarity between the above relation and
(\ref{eq:corrs'}), valid in the stationary scattering setup,
induces us to look for a more definite connection between the two
cases.

If we slightly generalize the abstract Hamiltonian 
(\ref{eq:modelH}) 
\andy{model2H}
\beq
H_{\rm dyn}=g[1+\alpha\tau_1+\beta\sigma_1+\gamma\tau_1\sigma_1],
\label{eq:model2H}
\eeq
by introducing the additional parameter $\gamma$, we easily find
the correspondence existing between the parameters involved: The
incident wave number $k$ of the neutron and the configuration of
the static potential (strength $B$ and width $a$) determine the
scattering data, which are reproducible by an appropriate choice of
parameters $\alpha,\beta,\gamma$ and $gT$ in the dynamical process
governed by the Hamiltonian (\ref{eq:model2H}).

For definiteness, consider the case of narrow potential, that is,
$a\to0$ or $ka\ll1$. Incidentally, notice that this is the case of
interest for the analysis of the QZE. The above $\xi_\pm$ and
$\phi_\pm$ are then approximated as
\andy{xiphi}
\beq
\xi_\pm\sim\pm\zeta ka,\qquad
\phi_\pm\sim(1\mp\zeta)ka,
\label{eq:xiphi}
\eeq
where we set $\zeta=\mu B/2E=m\mu B/k^2$, as in
Sec.~\ref{sec-numan}. In the limit $a\to0$, the evolution time $T$
is also considered to be of the same order of $a$ and the
transmission and reflection coefficients are expressed, in terms of
the parameters $\alpha,\beta,\gamma$ and $gT$, as
\andy{abgg'}
\beq
\left(\matrix{t_\uu\cr
              t_\dd\cr
              r_\uu\cr
              r_\dd\cr}\right)
\sim\left(\matrix{1\cr
               -i\beta gT\cr
               -i\alpha gT\cr
               -i\gamma gT\cr}\right).
\label{eq:abgg'}
\eeq
In the stationary scattering problem, the same quantities are calculated
to be
\andy{kxiphi}
\beq
\left(\matrix{t_\uu\cr
              t_\dd\cr
              r_\uu\cr
              r_\dd\cr}\right)
=\left(\matrix{e^{-ika}R'_{1,\u}\cr
               e^{-ika}R'_{1,\d}\cr
               L_{0,\u}\cr
               L_{0,\d}\cr}\right)
\sim\left(\matrix{1-ika+i(\phi_++\phi_-)/2\cr
               i(\phi_+-\phi_-)/2\cr
               -i(\xi_++\xi_-)/2\cr
               -i(\xi_+-\xi_-)/2\cr}\right)
\sim\left(\matrix{1\cr
                  -i\zeta ka\cr
                  0\cr
                 -i\zeta ka\cr}\right).
\label{eq:kxiphi}
\eeq
Therefore, the following abstract Hamiltonian
\andy{rH}
\beq
H_{\rm dyn}=\mu B(1+\tau_1)\sigma_1
\label{eq:rH}
\eeq
can reproduce the desired scattering data when the system evolves under
this Hamiltonian for time $T=a/v=ma/k$.

It is also interesting to see how such a dynamical Hamiltonian
$H_{\rm dyn}$ may reproduce the transfer matrix $M_\pm$
(\ref{eq:Mpm}), which further confirms the equivalence between the
two formalisms, stationary and dynamical, governed by the
Hamiltonians $H_{\rm Z}$ and $H_{\rm dyn}$, respectively. For this
purpose, consider first a neutron, initially prepared in state
$|R\pm\rangle$, subject to the dynamical evolution engendered by
$H_{\rm dyn}$ for time $T=ma/k$. By definition, the transfer matrix
connects the scattering products in the following way
\andy{Rinc}
\beq
\pmatrix{R_{1,\pm}^\prime \cr 0 \cr}
=M_\pm\pmatrix{1 \cr L_{0,\pm} \cr}.
\label{eq:Rinc}
\eeq
These scattering amplitudes are given by the corresponding
matrix elements of the evolution operator $e^{-iHT}$,
\andy{R'L}
\beq
e^{-ika}R_{1,\pm}'=\langle R\pm|e^{-iHT}|R\pm\rangle,\qquad
L_{0,\pm}=\langle L\pm|e^{-iHT}|R\pm\rangle,
\label{eq:R'L}
\eeq
which reduces, for small $T$, to
\andy{r'l}
\beq
R_{1,\pm}'\sim1+ika\mp i\mu BT,\qquad L_{0,\pm}\sim\mp i\mu BT.
\label{eq:r'l}
\eeq
On the other hand, if a neutron is prepared in $|L\pm\rangle$, we have
the relation
\andy{Linc}
\beq
\pmatrix{R_{1,\pm}' \cr e^{-ika} \cr}
=M_\pm\pmatrix{0 \cr L_{0,\pm} \cr},
\label{eq:Linc}
\eeq
where
\andy{R'Lr'l}
\beq
R_{1,\pm}'=e^{ika}\langle R\pm|e^{-iHT}|L\pm\rangle\sim\mp i\mu BT,\qquad
L_{0,\pm}=\langle L\pm|e^{-iHT}|L\pm\rangle\sim1\mp i\mu BT.
\label{eq:R'Lr'l}
\eeq
It is now an easy task to determine the matrix elements of $M_\pm$ from
the above relations (\ref{eq:Rinc})--(\ref{eq:R'Lr'l}).
We obtain
\andy{mpm}
\beq
M_\pm\sim\pmatrix{
1+ika\mp i\mu BT &\mp i\mu BT\cr
\pm i\mu BT      &1-ika\pm i\mu BT}
=1-i[\pm\mu B(i\tau_2+\tau_3)-2E\tau_3]T.
\label{eq:mpm}
\eeq
By defining a ``generator" $G_{\rm d}$
\andy{Gd}
\beq
G_{\rm d}=\mu B(i\tau_2+\tau_3)\sigma_1-2E\tau_3,
\label{eq:Gd}
\eeq
the transfer matrix $M_\pm$ for finite $a$ (or $T$) can be rewritten as
\andy{e-iGdT}
\beq
M_\pm
=\langle\pm|e^{-iG_{\rm d}T}|\pm\rangle,
\label{eq:e-iGdT}
\eeq
which is nothing but the transfer matrix (\ref{eq:Mpm}), obtained
for the stationary-state problem from the Hamiltonian $H_{\rm Z}$.



\begin{thebibliography}{99}

\bibitem{Beskow} \andy{Beskow}
A. Beskow and J. Nilsson, Arkiv f\"ur Fysik {\bf 34} (1967) 561;
L.A. Khalfin, Zh.\ Eksp.\ Teor.\
Fiz.\ Pis.\ Red.\ {\bf 8} (1968) 106 [JETP Letters {\bf 8} (1968) 65];
Phys.\ Lett.\  {\bf 112B} (1982) 223; Usp.\ Fiz.\ Nauk {\bf 160} (1990) 185
[Sov.\  Phys.\ Usp.\  {\bf 33} (1990) 10];
L. Fonda, G.C. Ghirardi, A. Rimini and T. Weber,
Nuovo Cim.\ {\bf A15} (1973) 689; {\bf A18} (1973) 805;
A. De Gasperis, L. Fonda
and G.C. Ghirardi, Nuovo Cim.\ {\bf A21} (1974) 471.

\bibitem{Misra} \andy{Misra}
B. Misra and E.C.G. Sudarshan, J. Math. Phys. {\bf 18} (1977) 756.

\bibitem{strev} \andy{strev} For a review on the temporal
behavior of quantum systems and the quantum Zeno effect, see
L. Fonda, G.C. Ghirardi and A.
Rimini, Rep.\ Prog.\ Phys.\ {\bf 41} (1978) 587; G-C. Cho, H. Kasari and Y.
Yamaguchi, Prog.\ Theor.\ Phys.\ {\bf 90} (1993) 803; H. Nakazato, M. Namiki
and S. Pascazio, Int.\ J. Mod.\ Phys.\ {\bf B10} (1996) 247;
D. Home and M.A.B. Whitaker, Ann.\ Phys.\ {\bf 258} (1997) 237.

\bibitem{Wilkinson} \andy{Wilkinson}
The first experimental observation of non-exponential decay at 
short times for an unstable system was performed two years ago: 
S.R. Wilkinson {\em et al.}, Nature {\bf 387} (1997) 575. No 
attempt has yet been made to suppress decay by repeated 
measurements (quantum Zeno effect). 

\bibitem{Cook} \andy{Cook} R.J. Cook, Phys.\ Scr.\ {\bf T21} (1988) 49.

\bibitem{Itano} \andy{Itano} W.H. Itano, D.J. Heinzen, J.J. Bollinger and
D.J. Wineland, Phys.\ Rev.\ {\bf A41} (1990) 2295.

\bibitem{Itanodiscuss} \andy{Itanodiscuss}
T. Petrosky, S. Tasaki and I. Prigogine, Phys.\ Lett.\ {\bf A151}
(1990) 109; Physica {\bf A170} (1991) 306; A. Peres and A. Ron,
Phys.\ Rev.\ {\bf A42} (1990) 5720; L.E. Ballentine, Phys.\ Rev.\
{\bf A43} (1991) 5165; W.H. Itano, D.J. Heinzen, J.J. Bollinger and
D.J. Wineland, Phys.\ Rev.\ {\bf A43} (1991) 5168; V. Frerichs and
A. Schenzle, in {\it Foundations of Quantum Mechanics}, T.D. Black,
M.M. Nieto, H.S. Pilloff, M.O. Scully and R.M. Sinclair, eds.
(World Scientific, Singapore, 1992); S. Inagaki, M. Namiki and T.
Tajiri, Phys.\ Lett.\ {\bf A166} (1992) 5; D. Home and M.A.B.
Whitaker, J. Phys.\ {\bf A25} (1992) 657; Phys.\ Lett.\ {\bf A173}
(1993) 327; Ph. Blanchard and A. Jadczyk, Phys.\ Lett.\ {\bf A183}
(1993) 272; T.P. Altenmuller and A. Schenzle, Phys.\ Rev.\ {\bf
A49} (1994) 2016; L.S. Schulman, A. Ranfagni and D. Mugnai, Phys.\
Scr.\ {\bf 49} (1994) 536; M. Berry, in {\it Fundamental Problems
in Quantum Theory}, eds., D.M. Greenberger and A. Zeilinger ({\it
Ann.\ N.Y. Acad.\ Sci.} {\bf Vol.\ 755}, New York, 1995), p.~303;
A. Beige and G. Hegerfeldt, Phys.\ Rev.\ {\bf A53} (1996) 53; A.
Luis and J. Peri\v na, Phys.\ Rev.\ Lett.\ {\bf 76} (1996) 4340. H.
Nakazato, M. Namiki, S. Pascazio and H. Rauch, Phys.\ Lett.\ {\bf
A217} (1996) 203; L.S. Schulman, J. Phys. {\bf A30} (1997) L293;
Phys. Rev. {\bf A57} (1998) 1509. K. Thun and J. Peri\v na, Phys.\
Lett.\ {\bf A249} (1998) 363.

\bibitem{PNBR} \andy{PNBR} S. Pascazio, M. Namiki, G. Badurek and H. Rauch,
Phys.\ Lett.\ {\bf A179} (1993) 155;
S. Pascazio and M. Namiki, Phys.\ Rev.\ {\bf A50} (1994) 4582.

\bibitem{inn} \andy{inn} P. Kwiat, H. Weinfurter, T. Herzog, A. Zeilinger and
M. Kasevich, Phys.\ Rev.\ Lett.\ {\bf 74} (1995) 4763.

\bibitem{NNPR} \andy{NNPR} H. Nakazato, M. Namiki, S. Pascazio and H. Rauch,
Phys.\ Lett.\ {\bf A199} (1995) 27;
Z. Hradil, H. Nakazato, M. Namiki, S. Pascazio and H. Rauch,
Phys.\ Lett.\ {\bf A239} (1998) 333.

 \bibitem{von} \andy{von}
J. von Neumann,  {\em  Die Mathematische Grundlagen der
Quantenmechanik} (Springer, Berlin, 1932). [English translation by
E.T. Beyer: {\em Mathematical Foundation of Quantum Mechanics}
(Princeton University Press, Princeton, 1955)]. For the QZE, see in
particular p.\ 195 of the German edition (p.\ 366 of the English
translation).

\bibitem{Wigner} \andy{Wigner} E.P. Wigner,
Am.\ J. Phys.\ {\bf 31} (1963) 6.

\bibitem{MScomment} \andy{MScomment}
Misra and Sudarshan in \cite{Misra} considered an initial
``unstable" state $\rho_0$ and took $E$ to be the subspace of the
undecayed states. The expression ``survived" means in their case
``undecayed."

\bibitem{Peres} \andy{Peres}
A. Peres, Am.\ J. Phys.\ {\bf 48} (1980) 931.

\bibitem{Jerica} \andy{Jerica}
E. Jericha, C.J. Carlile, M. Jaekel and H. Rauch, 
Physica {\bf B234-236} (1997) 1066.

\bibitem{spinflip} \andy{spinflip}
B. Alefeld, G. Badurek and H. Rauch, Z. Phys.\ {\bf 41B} (1981) 231.

\bibitem{fluctB} \andy{fluctB}
H. Rauch, M. Suda and S. Pascazio, ``Decoherence, dephasing and
depolarization" Physica {\bf B}, in print; G. Badurek, H. Rauch, M.
Suda and H. Weinfurter, ``Identification of a coherent
superposition of spin-up and spin-down states in neutron spin-echo
systems" preprint 1999.


\end{thebibliography}
\end{document}